\documentclass[aps,amsmath,amssymb,floatfix,twocolumn]{revtex4}

\usepackage{bm}
\usepackage{graphicx}
\usepackage{amsbsy}
\usepackage{dcolumn}

\newcommand{\be}{\begin{equation}}
\newcommand{\ee}{\end{equation}}
\newcommand{\bea}{\begin{eqnarray}}
\newcommand{\eea}{\end{eqnarray}}
\newcommand{\eqnm}[1]{Eq.\ (\ref{#1})}
\newcommand{\fignm}[1]{Fig.\ \ref{#1}}

\begin{document}

\title{Microscopic optical potential from chiral nuclear forces}

\author{J.\ W.\ Holt$^1$, N.\ Kaiser$^2$, G.\ A.\ Miller$^1$ and W.\ Weise$^{2,3}$}
\affiliation{$^1$Physics Department, University of Washington, Seattle,
  Washington 98195, USA}
\affiliation{$^2$Physik Department, Technische Universit\"{a}t M\"{u}nchen,
    D-85747 Garching, Germany}
\affiliation{$^3$ECT$^{\, *}$, Villa Tambosi, I-38123 Villazzano (TN), Italy}

\begin{abstract}

The energy- and density-dependent single-particle potential for nucleons is constructed in 
a medium of infinite isospin-symmetric nuclear matter starting from realistic
nuclear interactions derived within the framework of chiral effective field theory.
The leading-order terms from both two- and three-nucleon forces give rise to real,
energy-independent contributions to the nucleon self-energy. The Hartree-Fock contribution 
from the two-nucleon force is attractive and strongly momentum dependent, in contrast 
to the contribution from the three-nucleon force which provides a nearly constant 
repulsive mean field that grows approximately linearly with the nuclear density. 
Together, the leading-order perturbative contributions yield an attractive single-particle 
potential that is however too weak compared to phenomenology. Second-order
contributions from two- and three-body forces then provide the additional attraction required 
to reach the phenomenological depth. The imaginary part of the optical 
potential, which is positive (negative) for momenta below (above) the Fermi momentum, 
arises at second-order and is nearly inversion-symmetric about the Fermi
surface when two-nucleon interactions alone are present. The imaginary part is strongly 
absorptive and requires the inclusion of an effective mass correction as well as self-consistent
single-particle energies to attain qualitative agreement with phenomenology.

\end{abstract}

\maketitle
%\bigskip
%{\small PACS: 21.30.Fe, 21.60.Cs, 23.40.-s\\
%Keywords: Effective field theory at finite density, chiral three-nucleon
%force.}

%%%%%%%%%%%%%%%%%%%%%%%%%%%%%%%%%%%%%%%%
\section{Introduction}
%%%%%%%%%%%%%%%%%%%%%%%%%%%%%%%%%%%%%%%%

Nuclear optical model potentials provide a highly successful framework for describing 
nucleon-nucleus scattering across extended regions of the nuclear chart. While local and 
global phenomenological optical potentials \cite{becchetti69,varner91,koning03} have 
been used to describe total cross sections, elastic scattering angular distributions, and 
analyzing powers for reactions on target nuclei close to the valley of 
stability, microscopic optical potentials have no adjustable parameters and may 
therefore provide the best means for extrapolating to rare isotope reactions
that will be studied at the next generation of radioactive beam 
facilities. Neutron-capture cross sections on exotic, neutron-rich isotopes are 
particularly relevant for a detailed understanding of heavy-element formation 
in $r$-process nucleosynthesis. Although such reactions are experimentally 
unfeasible in the near future, neutron capture on rare isotopes can be 
probed indirectly in current and future rare isotope experiments through the 
$(d,p)$ stripping reaction, a process that is most easily modeled as a three-body 
problem requiring the nucleon-nucleon potential as well as the nucleon-nucleus 
optical potential \cite{nunes11} as input.

Phenomenological optical potentials possess several adjustable parameters 
that characterize the shape of the nuclear density distribution of the target nucleus
and that vary smoothly with the energy of the projectile and mass number of the target.
Microscopic optical potentials, on the other hand, are derived from an underlying 
model of the nuclear interaction fit to elastic nucleon-nucleon scattering data as well
as properties of the lightest nuclei. Within such a microscopic treatment, the optical 
potential is identified with the nucleon self-energy, a density-dependent complex-valued 
function 
given in terms of the nucleon energy and momentum. The nucleon self-energy 
has been constructed within numerous theoretical
frameworks, including Brueckner-Hartree-Fock (BHF) theory 
\cite{jeukenne76,grange87,haider88,bauge98,bauge01,hemalatha07,pachouri12}, 
Dirac-Brueckner-Hartree-Fock (DBHF) theory
\cite{arnold81,haar87,hama90,li92,xu12}, the Green's function formalism 
\cite{dickhoff04,waldecker11}, 
and chiral perturbation theory \cite{kaiser02,kaiser05}. The inclusion of three-nucleon forces, 
while often neglected in microscopic calculations of the optical potential, would seem 
highly relevant given their importance in achieving nuclear matter saturation at the
correct density and binding energy per particle. Nevertheless, recent BHF calculations
\cite{rafi13} included effects of the Urbana IX three-nucleon force \cite{pudliner95} 
in a simplified manner \cite{lejeune86} and found only a modest improvement in the 
comparison to elastic scattering data for intermediate-energy scattering of protons 
from $^{40}$C and $^{208}$Pb, despite a sizeable reduction of the central potential 
in the dense interior. A more accurate investigation of three-body forces is, however, 
desirable.

In the present work we make use of the progress that has been achieved in the 
last decade in constructing high-precision nuclear interactions within the framework 
of chiral effective field theory. As a first step in the development of microscopic
optical potentials capable of describing reactions on rare isotopes, we compute
the first- and second-order perturbative contributions to the nucleon self-energy in 
a medium of isospin-symmetric nuclear matter employing realistic chiral two- and 
three-nucleon interactions. Extensions to finite nuclei and isospin asymmetric
systems relevant for reactions on nuclei far from the valley of stability will be presented 
in future work. The resulting optical potentials for infinite nuclear matter can
be benchmarked against properties of well-established phenomenological potentials,
such as their depth and energy dependence.

We will show that at nuclear matter saturation density $\rho_0 \simeq 0.16$\,fm$^{-3}$, 
the leading-order Hartree-Fock contributions from two- and three-nucleon forces are 
strongly competitive, with the two-body component significantly attractive and the 
three-body component mildly repulsive. Alone they would give rise to a mean field whose 
depth for a nucleon at vanishing energy (with respect to the Fermi energy) would be 
$U \simeq -26$\,MeV, much smaller 
than the empirical value of $U \simeq -52$\,MeV determined from phenomenological optical
model fits to reactions on heavy stable nuclei \cite{koning03}. Second-order perturbative contributions 
from two- and three-nucleon forces
yield considerable additional attraction of approximately 30\,MeV, leading to 
overall reasonable agreement with phenomenology. The imaginary part, however, 
turns out to be nearly twice as strong as phenomenological optical potentials at 
intermediate scattering energies when single-particle energies are not treated 
self-consistently.

The paper is organized as follows. In Section \ref{momp} we introduce the relevant 
formalism and make a connection between the in-medium nucleon self-energy and
the nucleon-nucleus optical potential. Explicit formulas are given without any simplifying
approximations for the first and second-order perturbative contributions in terms of a 
partial-wave decomposition of the nucleon-nucleon interaction. We present as well
the formulas for the Hartree-Fock contribution to the single-particle potential from the
N$^2$LO chiral three-nucleon force. Section \ref{res} presents the numerical 
results for the momentum-dependent self-energy associated with negative-energy
hole states as well as positive-energy particle states. The impact of second-order 
three-body forces is then studied by employing a 
density-dependent nucleon-nucleon potential constructed by summing one nucleon 
over the filled Fermi sea. Our results for the real and imaginary potential depths 
as well as their energy dependence is compared to those of phenomenological 
optical potentials fit to reactions on stable nuclei. We end with a summary and conclusions.

%%%%%%%%%%%%%%%%%%%%%%%%%%%%%%%%%%%%%%%%%%%
\section{Microscopic optical model potentials}
\label{momp}
%%%%%%%%%%%%%%%%%%%%%%%%%%%%%%%%%%%%%%%%%%%
\subsection{First- and second-order contributions from two-body forces}

In the nuclear optical model, the complicated many-body problem associated with the
elastic scattering of a nucleon off a target nucleus is replaced by the more practicable 
problem of a single nucleon scattering from an equivalent complex mean-field potential:
\be
V(\vec r, \vec r^{\, \prime};E) = U(\vec r, \vec r^{\, \prime};E) + i W(\vec r, \vec r^{\ \prime};E),
\label{omp}
\ee
which in general is both non-local and energy-dependent. The imaginary part in 
\eqnm{omp} 
accounts for the presence of inelastic scattering, which reduces the total 
reaction flux in the elastic scattering channel. The simplest phenomenological 
optical potentials are taken to be local and of Woods-Saxon form in both the real and 
complex components:
\bea
U(r;E) &=& \frac{-U_0(E)}{1+e^{(r-R_r)/a_r}}, \nonumber \\
W(r;E) &=& \frac{-W_0(E)}{1+e^{(r-R_i)/a_i}},
\label{pomp}
\eea
where the parameters $U_0(E), W_0(E), R_{r,i}$ and $a_{r,i}$ vary smoothly with the 
mass number $A$ of the nucleus and, in the case of the well-depth parameters $U_0$ 
and $W_0$, also the projectile energy $E$. 

\begin{center}
\begin{figure*}
\begin{center}
\includegraphics[height=3.3cm]{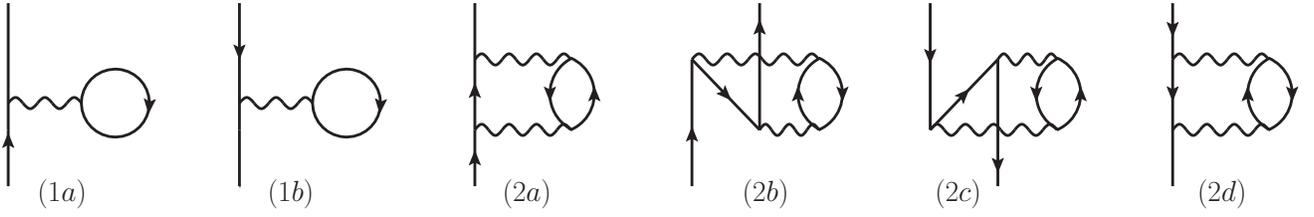}
\end{center}
\vspace{-.5cm}
\caption{Diagrams contributing to the nucleon self energy $\Sigma(q,\omega;k_f)$ at first and second order in 
perturbation theory from two-body forces. The first-order Hartree-Fock contributions are 
labeled (1a) and (1b) for particles $(q>k_f)$ and holes $(q<kf)$, respectively. The 
second-order contributions to the particle self-energy are labeled (2a) and (2b), while
the two contributions to the hole self-energy are labeled (2c) and (2d). The wavy line represents the 
{\it antisymmetrized} two-nucleon interaction $\bar V_{2N}$, including {\it direct and exchange} terms.}
\label{op2nf}
\end{figure*}
\end{center}

Beyond energies of $E \sim 200$\,MeV, this Woods-Saxon form is no longer sufficient, and
the real part of the central potential develops a ``wine-bottle'' shape \cite{koning03}.
Although not relevant for the present calculations with isospin-symmetric nuclear matter, 
phenomenological optical potentials possess real and imaginary spin-orbit terms 
as well as an imaginary surface term, all of which are proportional to the gradient of the
Woods-Saxon distribution. Extensive
analysis of the available experimental scattering data yields a real potential well
depth $U_0 \simeq 50-52$\,MeV for projectile nucleons with very low energies incident on 
heavy target nuclei. The depth of the imaginary potential vanishes at the Fermi
energy and grows to typical values of $W_0 \simeq 10-12$\,MeV for projectile
energies close to $100$\,MeV.

Microscopically the optical model potential can be identified with the nucleon self-energy
$\Sigma(\vec r, \vec r^{\, \prime};E)$ in a nucleus \cite{bell59}. 
For scattering states with $E>0$, $\Sigma(\vec r, \vec r^{\, \prime}, E)$ is the nuclear optical 
potential, while for bound states with $E<0$, the real part of $\Sigma(\vec r, \vec r^{\, \prime}, E)$ 
represents the shell model potential. In the present work we consider isospin-symmetric nuclear 
matter at uniform density $\rho = 2k_f^3/3\pi^2$, in which 
case it is more appropriate to compute the resulting spin- and isospin-independent self-energy 
in momentum-space $\Sigma(q,\omega;k_f)$. A local optical model potential for nucleon-nucleus 
scattering can then be obtained by solving the self-consistent equation for the on-shell energy 
in terms of the momentum and then folding the resulting density-dependent mean field with a 
realistic point-nucleon density distribution of the target nucleus. The off-shell dependence of 
the self-energy $\Sigma(q,\omega;k_f)$ on both  $q$ and $\omega$ is necessary to describe the nucleon spectral function and 
nucleon momentum distribution. A complementary work studying the off-shell self-energy, 
including the effects of three-nucleon forces, is given in Ref.\ \cite{wang13}.

The first-order Hartree-Fock contribution $\Sigma^{(1)}(q,\omega;k_f)$ to the self-energy from 
two-body forces is shown diagrammatically in \fignm{op2nf} for states above (1a) and 
below (1b) the Fermi surface. The Hartree-Fock contribution
\be
\Sigma^{(1)}_{2N}(q,\omega;k_f) = \sum_{1} \langle \vec q \, \vec h_1 s s_1 t t_1 | \bar V_{2N} | \vec q \,
\vec h_1 s s_1 t t_1 \rangle n_1,
\label{se1}
\ee
is real, $\omega$-independent, and changes smoothly as the external momentum $q$ crosses 
the Fermi surface. In Eq.\ (\ref{se1}), $\bar V_{2N}$ denotes the antisymmetrized potential, 
$n_1 = \theta(k_f-|\vec h_1|)$ is the 
zero-temperature occupation probability, and the sum is taken over the 
momentum, spin, and isospin of the intermediate hole state $|\vec h_1, s_1, t_1\rangle$.
The decomposition 
of the Hartree-Fock contribution in terms of partial-wave matrix elements of the interaction can be 
simplified by noting that $\Sigma(q,\omega;k_f)$ is spin and isospin independent when computed
for a background medium of isospin-symmetric nuclear matter. Averaging over $s$ and $t$ in 
Eq.\ (\ref{se1}) then yields the single-particle potential
\bea
&&\hspace{-.15in}U(q,k_f) = \frac{1}{2\pi^2}\sum_{lSJT} (2T+1)(2J+1) \nonumber \\
&&\hspace{-.15in}\times  \int^{(q+k_f)/2}_{{\rm max} \{0, (q-k_f)/2\} } \! dp \, p^2\, {\rm min} \{ 
2,(k_f^2-(q-2p)^2)/4pq \} \nonumber \\
&&\hspace{-.15in}\times \langle plSJT | \bar V_{2N} | plSJT \rangle,
\label{spen}
\eea
where $\vec p = (\vec q - \vec h_1)/2$ is the relative momentum of the interacting particles.

At second-order in perturbation theory, $\Sigma(q,\omega;k_f)$ develops both a real and imaginary part. 
For particle states above the Fermi surface, there are two distinct contributions 
labeled (2a) and (2b) in \fignm{op2nf}. The contribution (2a) arises from the external
particle coupling to a hole state inside the Fermi sea and reads:
\bea
&&\hspace{-.3in}\Sigma^{(2a)}_{2N}(q,\omega;k_f)  \nonumber \\
&&\hspace{-.15in}= \frac{1}{2}\sum_{123} \frac{| \langle \vec p_1 \vec p_3 s_1 s_3 t_1 
t_3 | \bar V_{2N} | \vec q \, \vec h_2 s s_2 t t_2 \rangle |^2}{\omega + \epsilon_2 - \epsilon_1
-\epsilon_3 + i \eta} \bar n_1 n_2 \bar n_3 \nonumber \\
&&\times (2\pi)^3 \delta(\vec p_1 + \vec p_3 - \vec q - \vec h_2),
\label{op2ac}
\eea
where $\bar n_k = 1-n_k$ denotes a particle state lying above the Fermi momentum.
We construct the momentum-dependent mean field by setting $\omega = q^2/(2M_N)$. 
Fixing $\vec p_3$ by momentum conservation, aligning the total
momentum $\vec p^{\, \prime} = \vec p_1 + \vec p_3 = \vec q + \vec h_2$ in the $\vec e_z$ 
direction, and averaging over the external particle spin, isospin and momentum
direction then yields the partial-wave decomposition:
\begin{widetext}
\bea
&& \hspace{-.1in}U(q,k_f) + i W(q,k_f) = \frac{8M_N}{(4\pi)^4q}
\sum_{\stackrel{l_1 l_2 l_3 l_4 J J^\prime M }{S m_s m_s^\prime T}}(2T+1)
\int_{p^\prime_a}^{p^\prime_b} dp^\prime 
\int_{q_{1a}}^{q_{1b}} d q_1\left [\int_0^{x_0} d\!\cos \theta_1 \bar 
P_{l_1,m}(\cos \theta_1)\bar P_{l_3,m}(\cos \theta_1)\right ] \nonumber \\
&& \times \int_{q_{2a}}^{q_{2b}} dq_2 \,
\bar P_{l_2,m^\prime}(\cos \theta_2)\bar P_{l_4,m^\prime}(\cos \theta_2)
\, \frac{p^\prime q_1^2 q_2}{(q_2-q_1+i\eta)(q_2+q_1)} i^{l_2+l_3-l_1-l_4} 
 \nonumber \\
&&\times  {\cal C}_{l_1mSm_s}^{JM}{\cal C}_{l_2m^\prime Sm_s^\prime}^{JM}
{\cal C}_{l_3mSm_s}^{J^\prime M}{\cal C}_{l_4m^\prime Sm_s^\prime}^{J^\prime M}
\langle q_1 l_1 S J T | \bar V_{2N} | q_2 l_2 S J T \rangle
\langle q_2 l_4 S J^\prime T | \bar V_{2N} | q_1 l_3 S J^\prime T \rangle,
\label{op2a}
\eea %remember 4pi convention here!
\end{widetext}
where $\vec q_1 = (\vec p_1 - \vec p_3)/2$, $\vec q_2 = (\vec q - \vec h_2)/2$,
$\bar P_{l m}$ is the associated Legendre function $P_{lm}$ multiplied by the factor
$\sqrt{(2l+1)(l-m)!/(l+m)!}$, $\cos \theta_2 = (q^2-q_2^2-{p^\prime}^2/4)/(p^\prime q_2)$, 
$x_0 = {\rm min}\{1,(q_1^2-k_f^2+{p^\prime}^2/4)/(p^\prime q_1)\}$, 
and the limits of integration are
\bea
&& p^\prime_a = {\rm max}\{0,q-k_f\}, \, p^\prime_b = q+k_f, \nonumber \\
&& q_{1a} = \sqrt{{\rm max}\{0,k_f^2-{p^{\prime}}^2/4\}}, \, q_{1b}=\infty \nonumber \\
&& q_{2a} = |q-p^\prime/2|, \nonumber \\
&& q_{2b} = {\rm min}\left \{\sqrt{(k_f^2+q^2)/2-{p^\prime}^2/4},q+p^\prime/2 \right \}.
\eea
The expression in 
Eq.\ (\ref{op2a}) holds also for the hole contribution labeled (2c) in \fignm{op2nf},
except that since $q<k_f$ the contribution is purely real and one can drop the 
$+i\eta$ in the energy denominator.

The diagrams labeled (2b) and (2d) in \fignm{op2nf} are both given by the following 
expression
\bea
&&\hspace{-.3in}\Sigma^{(2b)}_{2N}(q,\omega;k_f)  \nonumber \\
&&\hspace{-.15in}= \frac{1}{2}\sum_{123} \frac{| \langle \vec h_1 \vec h_3 s_1 s_3 t_1 
t_3 | \bar V_{2N} | \vec q \, \vec p_2 s s_2 t t_2 \rangle |^2}{\omega + \epsilon_2 - \epsilon_1
- \epsilon_3 - i \eta} n_1 \bar n_2 n_3 \nonumber \\
&&\times (2\pi)^3 \delta(\vec h_1 + \vec h_3 - \vec q - \vec p_2).
\label{op2bd}
\eea
In contrast to Eq.\ (\ref{op2ac}), here the contribution picks up an imaginary part for hole states
below the Fermi surface and is purely real for particle states above the Fermi surface. 
The partial-wave decomposition is very similar to that for $\Sigma^{(2a)}_{2N}(q,\omega;k_f)$, except
that $\vec q_1 = (\vec h_1 - \vec h_3)/2$, $\vec q_2 = (\vec q - \vec p_2)/2$, and one must
make the following replacements:
\bea
&&\hspace{-.15in}+i\eta \rightarrow -i\eta, \,\, x_0 \rightarrow {\rm min}\{1,(k_f^2-q_1^2-{p^\prime}^2/4)/(p^\prime q_1)\},  \nonumber \\
&& \hspace{-.15in}p^\prime_a \rightarrow {\rm max}\{0,k_f-q\}, \,\, p^\prime_b \rightarrow 2k_f, \nonumber \\
&& \hspace{-.15in}q_{1a} \rightarrow 0, \,\, q_{1b} \rightarrow \sqrt{k_f^2-{p^\prime}^2/4}, \nonumber \\
&&\hspace{-.15in} q_{2a} \rightarrow {\rm max}\left \{|q-p^\prime/2|, \sqrt{(k_f^2+q^2)/2-{p^\prime}^2/4} 
\right \}, \nonumber \\
&& \hspace{-.15in}q_{2b} \rightarrow q+p^\prime/2.
\eea

The numerical accuracy of the above formulas for the second-order contributions to
the nucleon self energy in nuclear matter has been checked against semi-analytic
expressions obtained for a simple scalar-isoscalar exchange model of the nuclear 
force (see the Appendix for details). Although the contributions labeled (2a) and (2c) 
in \fignm{op2nf} may potentially be divergent, for the scalar-isoscalar exchange interaction
all integrals converge. Across a range of momenta and densities we find the agreement 
between our numerical calculations and the semi-analytical results to be within 1\%. The formulas
for iterated one-pion exchange given in Refs.\ \cite{kaiser02,kaiser05} have been used as well
for checking the partial-wave representation of the second-order contribution.

%%%%%%%%%%%%%%%%%%%%%%%%%%%%%%%%%%%%%%%%%%%
\subsection{Leading-order contribution from three-body forces}

The methods described above for two-body forces can be extended to nuclear
many-body forces. For a general three-nucleon force, the first-order Hartree-Fock 
contribution to the nucleon self-energy is real and energy independent. Summing two
of the nucleons over the filled Fermi sea yields
\bea
&&\hspace{-.1in}\Sigma^{(1)}_{3N}(q,\omega;k_f) \\
&&= \sum_{12} \langle \vec q \, \vec h_1\vec h_2; s s_1s_2; t t_1t_2 |
 \bar V_{3N} | \vec q \, \vec h_1\vec h_2; s s_1s_2; t t_1t_2 \rangle n_1 n_2,
 \nonumber
\label{se31}
\eea
where $\bar V_{3N}$ is the fully-antisymmetrized three-body interaction.

In the present work we consider only the leading-order N$^2$LO chiral 
three-nucleon force, which has three terms proportional to 
the low-energy constants $c_1,c_3,c_4,c_D$, and $c_E$. The two-pion exchange
component has the momentum-space representation:
\be
V_{3N}^{(2\pi)} = \sum_{i\neq j\neq k} \frac{g_A^2}{8f_\pi^4} 
\frac{\vec{\sigma}_i \cdot \vec{q}_i \, \vec{\sigma}_j \cdot
\vec{q}_j}{(\vec{q_i}^2 + m_\pi^2)(\vec{q_j}^2+m_\pi^2)}
F_{ijk}^{\alpha \beta}\tau_i^\alpha \tau_j^\beta,
\label{3n1}
\ee
where $g_A=1.29$, $f_\pi = 92.4$ MeV, $m_{\pi} = 138$ MeV and $\vec{q}_i$ 
is the difference between the final and initial momenta of nucleon $i$. The isospin tensor 
\begin{equation}
F_{ijk}^{\alpha \beta} = \delta^{\alpha \beta}\left (-4c_1m_\pi^2
 + 2c_3 \vec{q}_i \cdot \vec{q}_j \right ) + 
c_4 \epsilon^{\alpha \beta \gamma} \tau_k^\gamma \vec{\sigma}_k
\cdot \left ( \vec{q}_i \times \vec{q}_j \right )
\label{3n4}
\end{equation}
results in two terms with the isospin structure $\vec \tau_i \cdot \vec \tau_j$ and one term 
proportional to $\vec \tau_k \cdot (\vec \tau_i \times \vec \tau_j)$. 
The one-pion exchange three-nucleon interaction is proportional to the low-energy 
constant $c_D$ and given by
\begin{equation}
V_{3N}^{(1\pi)} = -\sum_{i\neq j\neq k} \frac{g_A c_D}{8f_\pi^4 \Lambda_\chi} 
\frac{\vec{\sigma}_j \cdot \vec{q}_j}{\vec{q_j}^2+m_\pi^2} \vec{\sigma}_i \cdot
\vec{q}_j \, {\vec \tau}_i \cdot {\vec \tau}_j \, ,
\label{3n2}
\end{equation}
and finally the chiral three-nucleon contact interaction is proportional to the
low-energy constant $c_E$:
\begin{equation}
V_{3N}^{(\rm ct)} = \sum_{i\neq j\neq k} \frac{c_E}{2f_\pi^4 \Lambda_\chi}
{\vec \tau}_i \cdot {\vec \tau}_j\, ,
\label{3n3}
\end{equation}
where $\Lambda_{\chi} = 700$\,MeV sets the naturalness scale.

In the following, we will employ values of the low-energy constants
$c_1 =-0.81\,$GeV$^{-1}$, $c_3=-3.2\,$GeV$^{-1}$, and 
$c_4 =5.4\,$GeV$^{-1}$ for the two-pion exchange three-nucleon 
force, which can be constrained by nucleon-nucleon elastic scattering 
phase shifts \cite{entem03}. The low-energy constants $c_D$ and $c_E$ 
must be fit to nuclear systems with $A>2$. We employ the values
$c_D=-0.20$ and $c_E=-0.205$ extracted from a fit \cite{gazit09} to 
the binding energies of $A=3$ nuclei and the half-life of $^3$H. In fact,
the relevant dimensionful low-energy constants are $C_D = c_D/\Lambda_\chi$
and $C_E = c_E/\Lambda_\chi$ with values $C_D \simeq C_E \simeq -0.3$\,GeV$^{-1}$.

\begin{center}
\begin{figure*}
\begin{center}
\includegraphics[height=3.8cm]{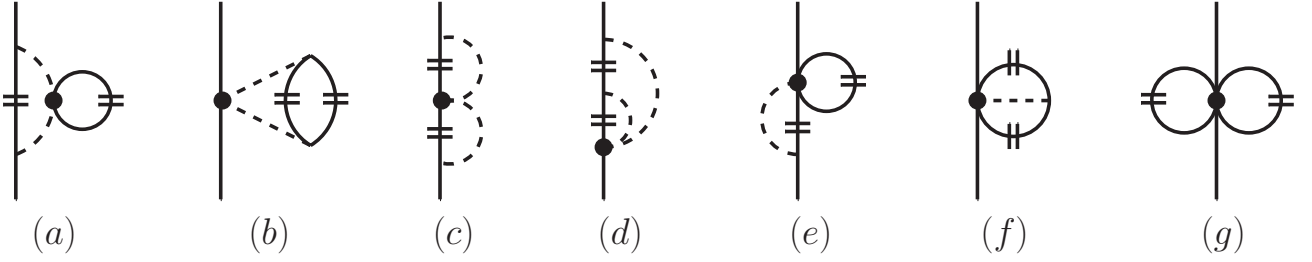}
\end{center}
\vspace{-.5cm}
\caption{Diagrammatic contributions from the N$^2$LO chiral three-nucleon force to the 
optical potential at first order in perturbation theory. The large dots represent vertices 
proportional to the low-energy constants $c_1,c_3,c_4,c_D,$ and $c_E$, while the short 
double-lines indicate a medium insertion $-2\pi \delta(k_0) \theta (k_f - |\vec k|)$. The 
external line can be either a hole or particle state. Reflected diagrams of (d) and (e) are 
not shown.}
\label{op3nf}
\end{figure*}
\end{center}

In \fignm{op3nf} we show the diagrammatic contributions to the nucleon
self-energy arising from the leading-order chiral three-nucleon force. The 
direct Hartree diagrams, labeled as (a) and (b) in \fignm{op3nf}, of the chiral 
two-pion exchange three-nucleon force are non-vanishing only for the terms 
proportional to the low-energy constants $c_1$ and $c_3$. 
The sum of these two diagrams gives
\bea
&&\hspace{-.15in}U(q,k_f) ={g_A^2 m_\pi^6\over (2\pi f_\pi)^4}\bigg\{
14(c_3-c_1)u^4+(3c_1-2c_3)u^2 \nonumber \\
&&\hspace{-.15in} -4c_3 u^6 +(12c_1-10c_3)u^3 \Big[\arctan 2u+\arctan(u+x) \nonumber \\
&&\hspace{-.15in} +\arctan(u-x)\Big]  + \Big[{c_3\over 2}(1+9u^2)-{3c_1\over 4}(1+8u^2)\Big] \nonumber \\ 
&&\hspace{-.15in} \times \ln(1+4u^2)+{u^3\over x}\Big[3c_3-4c_1+2(c_1-c_3)(x^2-u^2)\Big] \nonumber \\
&&\hspace{-.15in} \times \ln{1+(u+x)^2 \over 1+(u-x)^2}\bigg\}\,,
\label{hc134}
\eea
where $ u = k_f/m_\pi$ and $x = q/m_\pi$.

The Fock diagrams, labeled as (c) and (d) in \fignm{op3nf} are non-vanishing for all
terms in the two-pion exchange three-nucleon force:
\bea
&&\hspace{-.15in}U(q,k_f) ={g_A^2 m_\pi^6\over (4\pi f_\pi)^4x^2}\bigg\{3c_1 H^2(x,u)+
\Big({c_3\over 2}-c_4\Big)G_S^2(x,u) \nonumber \\
&&+(c_3+c_4)G_T^2(x,u) + \int_0^u\! d\xi\bigg[6c_1  H(\xi,u){\partial
 H(\xi,x)\over \partial x} \nonumber \\
&& +(c_3-2c_4)G_S(\xi,u){\partial G_S(\xi,x)\over \partial x} \nonumber \\
&&+2(c_3+c_4) G_T(\xi,u){\partial G_T(\xi,x)\over \partial x}\bigg]\bigg\}\,,
\eea
with the auxiliary functions:
\bea 
&&\hspace{-.40in}H(x,u) = u(1+x^2+u^2) \nonumber \\
&&\hspace{-.30in}-{1\over 4x}\Big[1+(u+x)^2\Big]\Big[1+(u-x)^2 \Big] \ln{1+(u+x)^2\over 1+(u-x)^2} \,,
\eea
\bea
&& \hspace{-.42in}G_S(x,u) = {4ux \over 3}( 2u^2-3) +4x\Big[\arctan(u+x)\nonumber \\
&&\hspace{-.30in}+\arctan(u-x)\Big] + (x^2-u^2-1) \ln{1+(u+x)^2\over  1+(u-x)^2} \, ,
\eea
\bea
&&\hspace{-.15in}G_T(x,u) = {ux\over 6}(8u^2+3x^2)-{u\over
2x} (1+u^2)^2  + {1\over 8} \bigg[ {(1+u^2)^3 \over x^2} \nonumber \\ 
&& -x^4+(1-3u^2)(1+u^2-x^2)\bigg] \ln{1+(u+x)^2\over  1+(u-x)^2} \,.
\eea
As shown in Section \ref{res}, the sum of the Hartree and Fock contributions from 
the two-pion exchange three-nucleon force gives rise to a significantly repulsive
mean field. The Hartree term is approximately 75\% larger in magnitude and of opposite 
sign as the attractive Fock term.

The contribution to the single-particle potential arising from the one-pion exchange
three-nucleon force, proportional to $c_D$, is given by
\bea
&&\hspace{-.20in}U(q,k_f) ={g_A c_D m_\pi^6\over (2\pi f_\pi)^4\Lambda_\chi}
\bigg\{u^6-{7u^4\over 4}+{u^2\over 8} \\
&&\hspace{-.15in} -{1+12u^2 \over 32} \ln(1+4u^2)+u^3\Big[\arctan 2u+\arctan(u+x) \nonumber \\
&&\hspace{-.15in}+\arctan(u-x)\Big] +
{u^3 \over 4x}(x^2-u^2-1)\ln{1+(u+x)^2 \over 1+(u-x)^2} \bigg\}\,,  \nonumber
\label{cdpart}
\eea
which depends very weakly on the momentum $q$ and is attractive for $c_D < 0$.
The first-order contribution from the N$^2$LO contact interaction is independent of the
external momentum and has the form
\begin{equation}
U(q,k_f) = -{c_E k_f^6 \over 4\pi^4 f_\pi^4 \Lambda_\chi} \,,
\label{ctnf}
\end{equation}
which is of course repulsive for $c_E<0$. As we will find in Section \ref{res}, 
together $V_{3N}^{1\pi}$ and $V_{3N}^{ct}$ provide a nearly constant repulsive mean field.

\begin{center}
\begin{figure*}[t]
\begin{center}
\includegraphics[height=18cm,angle=270]{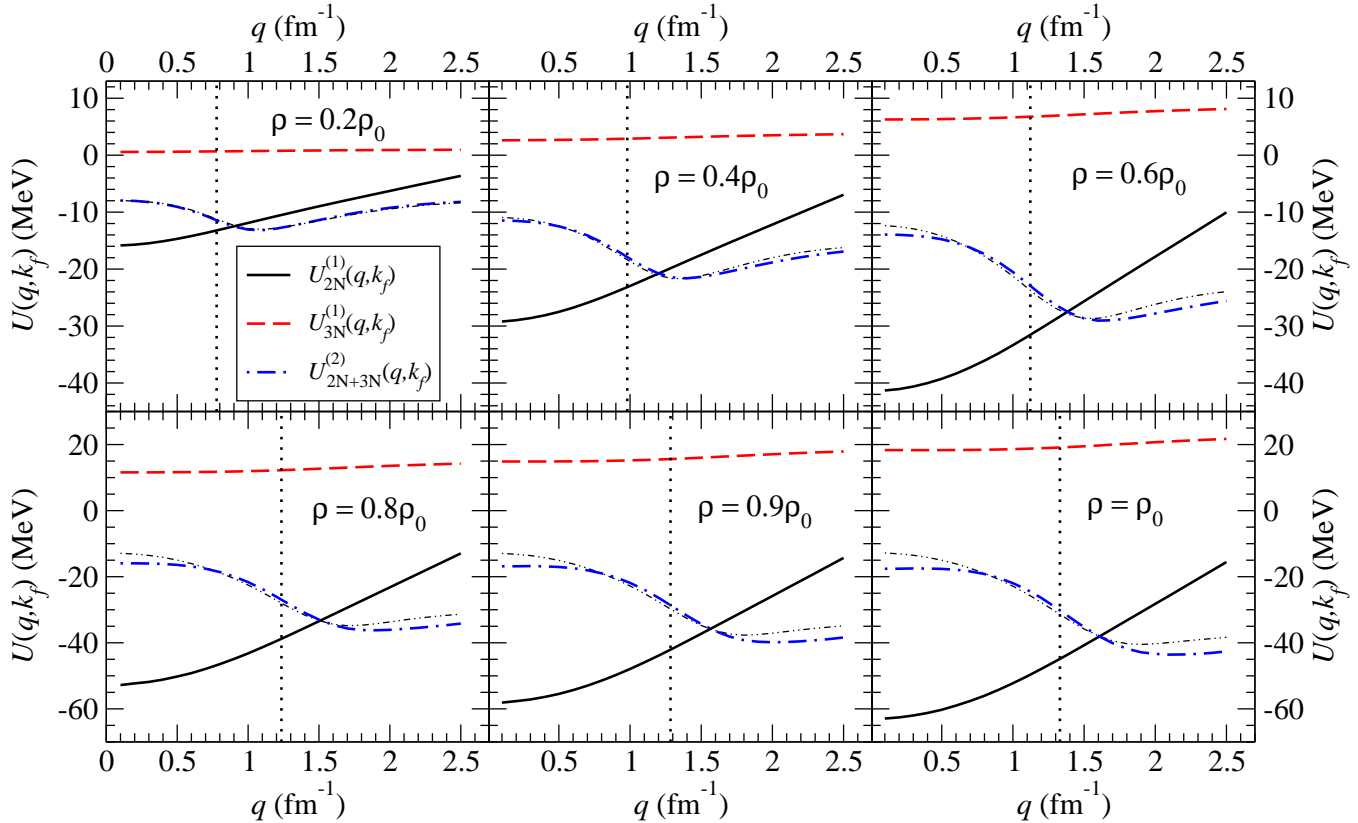}
\end{center}
\vspace{-.5cm}
\caption{Contributions to the real part of the momentum- and density-dependent optical potential. 
The solid and dashed-dotted lines are the first- and second-order contributions, respectively from 
the N$^3$LO chiral two-body potential, while the dashed line is the first-order contribution from 
the N$^2$LO chiral three-nucleon force. The vertical dotted line denotes the Fermi momentum, and 
the dashed-double-dotted line denotes the second-order contribution without three-body forces. The results are
shown for the case $\omega = q^2/(2M_N)$.}
\label{ddopr}
\end{figure*}
\end{center}

The above analytical expressions result from an exact calculation of the Hartree-Fock
contribution to the nuclear mean field. To include second-order corrections from three-nucleon
forces, we compute the expressions in Eqs.\ (\ref{op2ac}) and (\ref{op2bd}) using a 
density-dependent two-body effective interaction \cite{holt09,holt10,hebeler10}.

%%%%%%%%%%%%%%%%%%%%%%%%%%%%%%%%%%%%%%%%%%%%%%%
%%%%%%%%%%%%%%%%%%%%%%%%%%%%%%%%%%%%%%%%%%%%%%%
\section{Results}
\label{res}
%%%%%%%%%%%%%%%%%%%%%%%%%%%%%%%%%%%%%%%%%%%%%%%

In the present section we employ the N$^3$LO chiral two-body interaction of Ref.\ \cite{entem03} 
together with the N$^2$LO chiral three-body interaction with low-energy constants given in 
Section \ref{momp} to compute the contributions to the nuclear optical potential up to second 
order in perturbation theory. In addition we perform calculations of the nuclear mean field also
for hole states with $q<k_f$. We are particularly interested in comparisons of our microscopic
optical potential to local phenomenological potentials and in the effects from three-nucleon forces, 
which until now have been treated only approximately in several complementary studies 
\cite{rafi13,wang13}.

\begin{figure}
\includegraphics[height=8.7cm,angle=270]{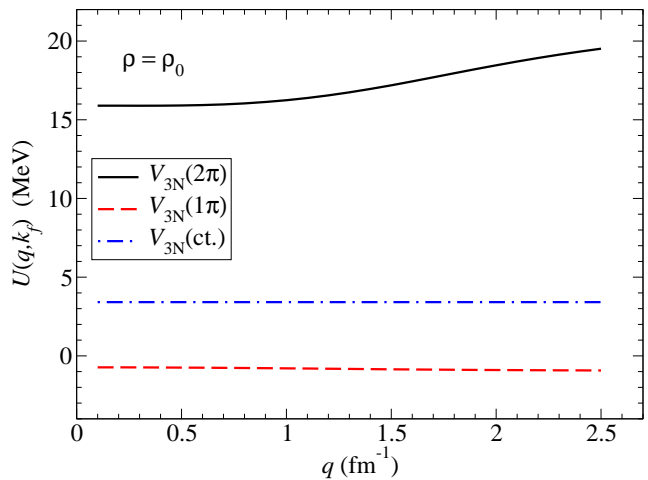}
\vspace{-.5cm}
\caption{Hartree-Fock contributions to the real part of the nuclear optical potential 
from chiral three nucleon forces. The two-pion exchange, one-pion exchange, and 
contact three-nucleon force contributions are evaluated from Eqs.\ (\ref{hc134})-(\ref{ctnf}) 
and plotted separately as a function of the momentum.}
\label{ddop3}
\end{figure}

In \fignm{ddopr} we plot the real part of the on-shell self-energy ($\omega = q^2/(2M_N)$) 
as a function of momentum and
density. The thick solid line denotes the Hartree-Fock contribution from two-body forces, and the
vertical dotted lines show the Fermi momentum corresponding to the densities 
$\rho = \{0.2\rho_0,0.4\rho_0,0.6\rho_0,0.8\rho_0,0.9\rho_0,\rho_0\}$ from the upper left 
corner to the bottom right, where $\rho_0 \simeq 0.16$\,fm$^{-3}$. The
Hartree-Fock term has a nearly parabolic form, and when summed with the free-particle kinetic 
energy $q^2/(2M_N)$ it can be well approximated as \cite{holt11}:
\be
\epsilon_q = \frac{q^2}{2M^*} + \Delta,
\ee
where $M^*$ is the effective mass and the energy shift $\Delta$ is independent of momentum. 
In \fignm{ddopr} the momenta are taken up to 
$q=2.5$\,fm$^{-1}$, which, for all the densities considered here, corresponds to possible two-particle
relative momenta well below the cutoff of $\Lambda \simeq 2.5$\,fm$^{-1}$.

\begin{figure}
\includegraphics[height=8.7cm,angle=270]{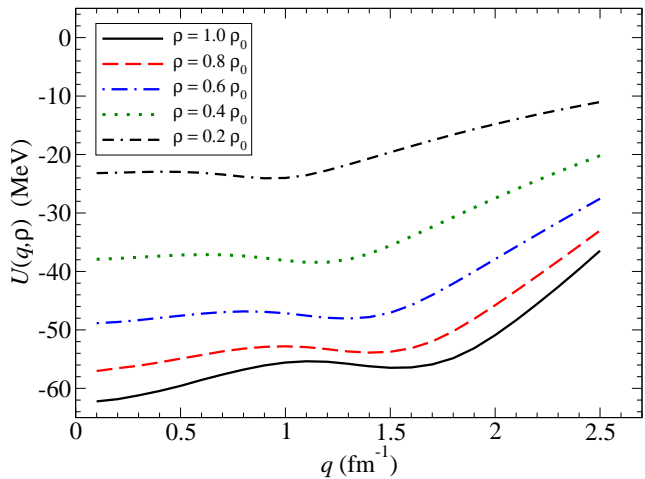}
\vspace{-.5cm}
\caption{The real part of the momentum-dependent optical potential at second
order in perturbation theory from chiral two- and three-nucleon forces. The optical
potential is computed for a medium of symmetric nuclear matter at densities 
ranging from $0.2\rho_0$ to $\rho_0$.}
\label{ddopa}
\end{figure}

The Hartree-Fock three-body force contribution exhibits a very weak density dependence that
would give rise to only a small decrease in the effective mass at the Fermi surface \cite{holt12}.
The strength of the mean field from chiral three-nucleon forces increases nearly linearly with
the density of the medium. At saturation density it gives a repulsive contribution of 
approximately $20$\,MeV. In \fignm{ddop3} we plot separately the mean fields associated with 
the different contributions $V_{3N}^{2\pi}$, $V_{3N}^{1\pi}$, and $V_{3N}^{ct}$ at nuclear matter 
saturation density, corresponding to $k_f=1.33$\,fm$^{-1}$. The $2\pi$-exchange chiral 
three-nucleon force provides much of the observed repulsion from three-body forces and accounts
also for most of the momentum dependence, which arises primarily for momenta above the Fermi
surface. The $1\pi$ and contact interactions together give rise to a small net repulsive mean field
that is nearly momentum independent. For nucleon-nucleus scattering, it therefore 
appears that the low-energy constants $c_D$ and $c_E$ are strongly correlated, with variations
along the line
\be
c_E = \alpha \cdot c_D + {\rm const}
\ee
giving nearly equivalent descriptions of the mean field, where the constant of proportionality 
$\alpha \simeq 0.21\pm 0.02$ is weakly dependent on momentum and density. Inspection of
Eq.\ (\ref{cdpart}) reveals that in the chiral limit only the leading $k_f^6$ term survives, and
the correlation coefficient would be $\alpha = g_A/4 \simeq 0.3$.

The second-order contributions to the nuclear mean field are shown as the dashed-dotted
lines in \fignm{ddopr}. Below the Fermi surface, they have a momentum dependence that is
nearly opposite to that of the Hartree-Fock contribution, giving rise to a quasiparticle effective 
mass at the Fermi surface that is close to the mass in vacuum \cite{holt11}. In 
\fignm{ddopr} we plot also the second-order contribution without three-nucleon forces, denoted 
by the dashed-double-dotted line. Despite the fact that the three-nucleon
force gives rise to substantial repulsion at the Hartree-Fock approximation, it appears that
second-order effects are quite small and produce additional attraction at both low and
high momenta. 

It is common in the literature to include self-consistent single-particle
energies in the denominators of the second-order contributions. Then the on-shell
condition reads:
\be
\epsilon_p = \frac{p^2}{2M_N} + {\rm Re}\, \Sigma(p,\epsilon_p;k_f).
\label{sce}
\ee
Such a prescription reduces the second-order contributions due to the larger energy difference
between particle and hole states. The value in using the free-particle spectrum is
that various thermodynamic identities, such as the Hugenholtz--Van-Hove \cite{hugenholtz58} 
and Luttinger \cite{luttinger61} theorems
\bea
&&\frac{k_f^2}{2M_N} + U(k_f,k_f) = \bar E(k_f) + \frac{k_f}{3}\frac{\partial \bar E(k_f)}{\partial k_f}\nonumber \\
&&W(q,k_f) = C\, |k_f-q| (k_f-q) + \cdots
\eea
are automatically fulfilled when the relevant quantities 
are computed to a particular order in perturbation theory. Nevertheless, to achieve a better 
description of phenomenology may require using the self-consistent energies \eqnm{sce}.

\begin{figure}
\includegraphics[height=8.7cm,angle=270]{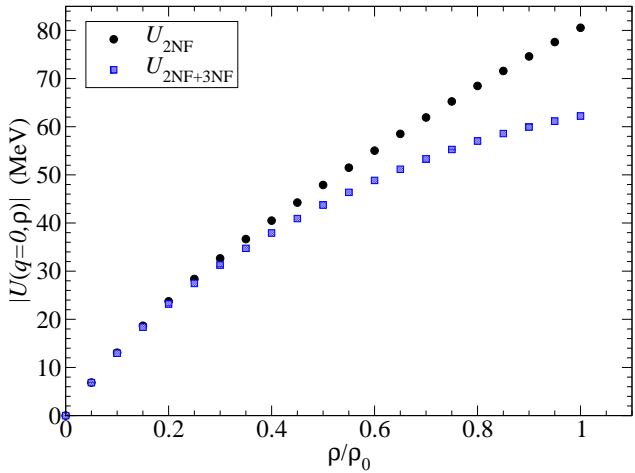}
\vspace{-.5cm}
\caption{Density dependence of the real part of the optical potential at zero momentum from
second order perturbation theory. Results for two-nucleon forces alone as well as for the sum
of two- and three-nucleon forces are shown.}
\label{ddopa2}
\end{figure}

The combined real part of the nucleon self-energy is shown in \fignm{ddopa} as a function of
density and momentum. We note that for low to moderate densities, the mean field for states with
momenta $q<k_f$ is nearly constant, but in the vicinity of the saturation density, three-nucleon forces
at second-order introduce additional attraction for low values of $q$. The well depth
for a scattering state at zero incident 
energy is approximately $-57$\,MeV, which is within 10\% of the depth, $-52$\,MeV, of 
phenomenological optical potentials. The well depth at $q=0$ as a function of density is shown in 
Fig.\ \ref{ddopa2} for two-nucleon forces alone as well as for combined two- and three-body
forces. Three nucleon forces become relevant at about 40\% of nuclear matter saturation
density and result in a mean field that is significantly nonlinear in the density.

\begin{center}
\begin{figure*}
\begin{center}
\includegraphics[height=18cm,angle=270]{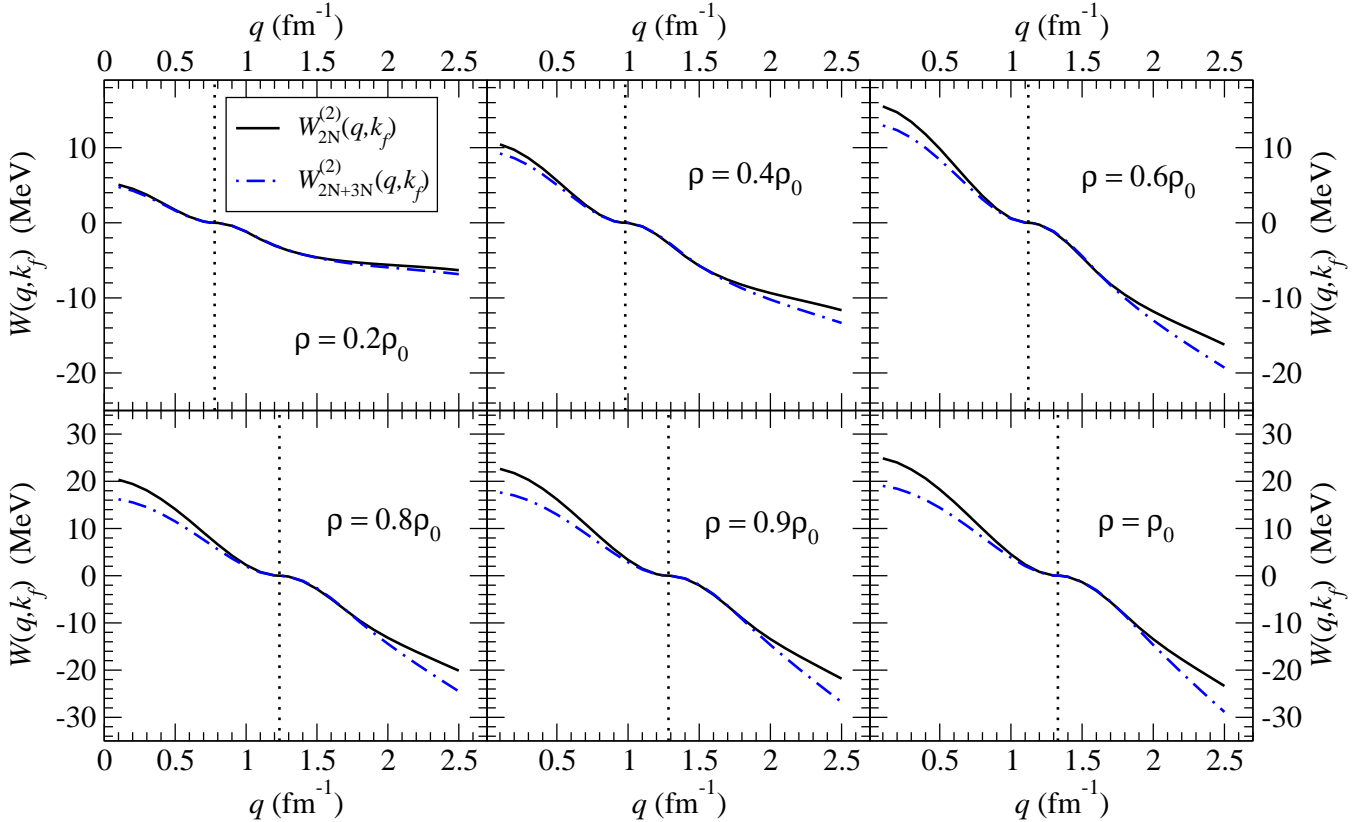}
\end{center}
\vspace{-.5cm}
\caption{The imaginary part of the momentum- and density-dependent optical potential arising
from chiral two- and three-nucleon forces iterated to second order. The vertical dotted line 
denotes the Fermi momentum.}
\label{ddopi}
\end{figure*}
\end{center}

Finally, we plot in \fignm{ddopi} the imaginary part of the nucleon self-energy arising from the
second-order perturbative contributions (both with and without three-nucleon forces) as a function
of momentum and density. In agreement with Luttinger's theorem \cite{luttinger61}, the imaginary 
part vanishes quadratically in the vicinity of the Fermi surface above and below $k_f$ for both
two- and three-nucleon force contributions. Omitting the chiral three-body force, we find that the 
imaginary part is approximately inversion-symmetric about the Fermi momentum, $W(q,k_f) 
\simeq -W(2k_f-q,k_f)$, a property which is often 
assumed in the dispersion optical model formalism \cite{mahaux86}. This feature is, however, 
modified with the inclusion of three-nucleon forces, which provide an attractive contribution 
at both very low and very high momenta. 

The overall strength of the imaginary potential at nuclear matter saturation density for an intermediate 
scattering energy of $E \simeq 100$\,MeV is approximately 
$30$\,MeV, which would seem too large compared to the empirical value,
$|W| \simeq 10$\,MeV \cite{koning03}. This large magnitude of the imaginary part of the optical potential
is a feature shared by many microscopic calculations. Already second-order one-pion exchange gives
rise to quantitatively similar results (see Fig.\ 6 in Ref.\ \cite{kaiser05}).
However, as noted in Refs.\ \cite{negele81,fantoni81}, the 
imaginary part of the self-energy should not be compared with the imaginary part of phenomenological
optical potentials. Rather, the precise relationship between the microscopic and phenomenological 
potential is given by
\be
W_{ph}(q,E) = \left (1+\frac{M_N}{q}\frac{\partial U}{\partial q}\right )^{-1} W_{mc}(q,E).
\ee
With the so-called $k$-mass factor $\left (1+\frac{M_N}{q}\frac{\partial U}{\partial q}\right )^{-1} \simeq 0.75$,
the value of the imaginary part of the optical potential derived from our microscopic calculation is
$W(q=100\,{\rm MeV}) = -22.5$\,MeV. Since this value is still rather large compared to the empirical
optical potential, it appears that the inclusion of self-consistent energies in the denominators of 
Eqs.\ (\ref{op2ac}) and (\ref{op2bd}) may be necessary in order to achieve a quantitatively successful 
imaginary microscopic potential. In addition one should recall that the phenomenological absorptive
strength $|W|$ is deduced for finite nuclei which have a characteristic gap in the single-particle 
energy spectrum around the Fermi energy. The nuclear matter calculation does not feature such a 
gap at the Fermi surface so that there is an increased phase space open for absorptive processes,
leading to an overestimate of $|W|$.

%%%%%%%%%%%%%%%%%%%%%%%%%%%%%%%%%%%%%%%%%%%%%
%%%%%%%%%%%%%%%%%%%%%%%%%%%%%%%%%%%%%%%%%%%%%
\section{Conclusions}
%%%%%%%%%%%%%%%%%%%%%%%%%%%%%%%%%%%%%%%%%%%%%

We have performed a microscopic calculation of the on-shell self-energy of a nucleon in
a medium of isospin-symmetric nuclear matter at uniform density $\rho$ up to second order in
many-body perturbation theory. The starting point is a realistic N$^3$LO chiral two-nucleon
potential supplemented with the N$^2$LO chiral three-nucleon force.  
The first- and second-order contributions 
from two-body forces are attractive, but below the Fermi momentum they have an opposite 
dependence on the momentum. The N$^2$LO chiral three-body force is found to provide substantial 
repulsion that grows slowly with momentum and nearly linearly with the density. 
Summing up all of these contributions, the resulting microscopic
nuclear mean field agrees qualitatively with the depth of phenomenological optical potentials. 
The absorptive strength of the imaginary
part of the potential calculated in nuclear matter is considerably larger than the empirical
one deduced for finite nuclei. This suggests that a 
self-consistent treatment of single-particle energies (including the energy gap at the Fermi surface)
may be necessary in order to achieve
a successful description of nucleon-nucleus scattering at low to intermediate energies.
In the future we plan to extend our calculations to finite nuclei and isospin
asymmetric nuclear matter that will be important to describe neutron-capture cross
sections on neutron-rich isotopes.

Work supported in part by BMBF, the DFG cluster of excellence Origin and 
Structure of the Universe, by the DFG, NSFC (CRC110) and US DOE Grant No.\ 
DE-FG02-97ER-41014.

%%%%%%%%%%%%%%%%%%%%%%%%%%%%%%%%%%%%%%%%%%%%%
%%%%%%%%%%%%%%%%%%%%%%%%%%%%%%%%%%%%%%%%%%%%%

\section{Appendix: Nuclear optical potential from
second-order scalar-isoscalar boson exchange}
\label{appen}
As a benchmark for our involved numerical calculations of the nuclear mean field
at second order in perturbation theory, we derive exact semi-analytical expressions
for the on-shell self-energy arising from scalar-isoscalar boson exchange.
The attractive central NN-potential in momentum space is given by:
\begin{equation} V_C(Q) = - {g^2 \over m^2+Q^2}\,, \end{equation}
with $g$ the coupling constant, $m$ the boson mass, and $Q$ the
momentum transfer between the two nucleons.

The first-order contribution to the real part of the optical potential for states both
above ($q>k_f)$ and below ($q<k_f)$ the Fermi surface reads:
\bea
&&U(q,k_f)^{(1)} = {g^2 m \over 4\pi^2} \bigg\{
- \arctan(u+x)-\arctan(u-x) \nonumber \\ 
&& +u -{8u^3 \over 3}
+{1+u^2-x^2 \over 4x} \ln{1+(u+x)^2 \over 1+(u-x)^2} \,,
\eea
with abbreviations $u= k_f/m$ and $x= q/m$.

Due to the presence of poles in Fermi sphere integrals, the analytic expression of the second-order 
contributions cannot be continued directly from below to above the Fermi surface. We 
therefore distinguish the contributions to the optical potential for momenta $q<k_f$ 
and $q>k_f$. Setting $\omega = q^2/(2M_N)$, the complex-valued mean field $U(q,k_f)+i\,W(q,k_f)$ 
inside the Fermi sphere $q<k_f$ is given by the sum of the following contributions
(in these expresssions the superscript ``$H$'' and ``$F$'' refer to Hartree and Fock diagrams, and the subscript
denotes the number of medium insertions \cite{kaiser02}):
\bea
&&\hspace{-.2in}U_2(q,k_f)^{(H)} = {g^4 M_N \over 8\pi^3} \bigg\{
\arctan(u+x)+\arctan(u-x) \nonumber \\
&&\hspace{-.2in} -u+{x^2-u^2-1 \over 4x}
\ln{1+(u+x)^2 \over 1+(u-x)^2} \,, 
\eea

\bea
&&\hspace{-.1in}U_2(q,k_f)^{(F)} = {g^4 M_N \over 16\pi^3}\bigg\{
\int_0^{(u-x)/2} d\xi \, 8\xi +  \int_{(u-x)/2}^{(u+x)/2} d\xi
 \nonumber \\
&& \hspace{-.1in} \times {1\over x} \Big[u^2 -(2\xi-x)^2\Big] \bigg\} 
{\arctan 2\xi-\arctan \xi \over 1+2\xi^2} \,,
\eea

\bea
&& \hspace{-.14in} U_3(q,k_f)^{(H)} = {g^4 M_N \over 8\pi^4}
\int_{-1}^1 dy\,\Bigg\{\bigg[ uxy +{1\over 2} (u^2-x^2y^2) \nonumber \\
&& \hspace{-.14in} \times \ln{u+x y \over u-x y} \bigg] {s^2 \over 1+s^2} 
+\int_{-xy}^{s-x y} d\xi \bigg[2u \xi +(u^2-\xi^2)  \ln{u+\xi \over u-\xi} \bigg] \nonumber \\
&& \hspace{-.14in} \times {x y +\xi \over [1+(xy +\xi)^2]^2}
+{1\over x} \int_0^u d\xi \, {\xi^2 \sigma^2\over 1+\sigma^2}
 \ln{|x+\xi y| \over |x-\xi y|} \Bigg\}\,, 
\eea
with auxiliary functions $s= x y +\sqrt{u^2-x^2+x^2y^2}$ and
$\sigma= \xi y +\sqrt{u^2-\xi^2+\xi^2y^2}$.

\bea
&& \hspace{-.17in} U_3(q,k_f)^{(F)} = {g^4 M_N \over 16\pi^4}
\int_{-1}^1 dy\, \Bigg\{\int_0^u d\xi\, {\xi^2 \over x R }\ln(1+\sigma^2) \nonumber \\
&& \hspace{-.17in}  \times \ln{|x R +(x^2-\xi^2-1) y \xi| \over |x R +(1+\xi^2-x^2) y \xi|} 
-\int_{-1}^1dz\,{ y z \,\theta(y^2+z^2-1)
\over 4|y z|\sqrt{y^2+z^2-1}}\, \nonumber \\ 
&& \hspace{-.17in}  \times \ln(1+s^2)\ln(1+t^2) 
\Bigg\}\,, 
\eea
with auxiliary functions $t= x z +\sqrt{u^2-x^2+x^2z^2}$ and $R = \sqrt{(1+x^2-\xi^2)^2
+4\xi^2(1-y^2)}$.

\bea
&&\hspace{-.45in} W_2(q,k_f)^{(H)} = {g^4 M_N \over 16\pi^3} \bigg\{
\ln[1+(u+x)^2] \nonumber \\
&&\hspace{-.45in}+\ln[1+(u-x)^2] -2(1+u^2) +{2x^2 \over 3} \nonumber \\
&&\hspace{-.45in}+{1+u^2-x^2 \over x} \Big[
\arctan(u+x)-\arctan(u-x)\Big] \bigg\} \,,
\eea

\bea
&& \hspace{-.45in} W_2(q,k_f)^{(F)} = {g^4 M_N \over 32\pi^3}\bigg\{
\int_0^{(u-x)/2} d\xi \, 8\xi \nonumber \\
&& \hspace{-.45in}+ \int_{(u-x)/2}^{(u+x)/2} d\xi \,
{1\over x}\Big[u^2-(2\xi-x)^2\Big] \bigg\} {\ln(1+4\xi^2) \over
1+2\xi^2} \,, 
\eea

\bea
&& \hspace{-.15in} W_3(q,k_f)^{(H)} = {g^4 M_N \over 16\pi^3}
\int_{-1}^1 dy\,\Bigg\{(1+2u^2-2x^2y^2){s^2 \over 1+s^2} \nonumber \\
&& \hspace{-.15in} -\ln(1+s^2) +2x y\bigg( \arctan s -{s\over 1+s^2} \bigg) \nonumber \\
&& \hspace{-.15in} +\int_0^u d\xi \, {2\xi^2\over x}\,\theta(x-\xi|y|) {\sigma^2\over
1+\sigma^2} \Bigg\}\,,
\eea

\bea
&& W_3(q,k_f)^{(F)} = {g^4 M_N \over 16\pi^3}
\int_{-1}^1 dy\, \Bigg\{ -\int_{-1}^1dz\, \ln(1+s^2) \nonumber \\
&& \times \ln(1+t^2) {\theta(1-y^2-z^2) \over 4\pi \sqrt{1-y^2-z^2}}\, \nonumber \\
&& -\int_0^u d\xi\, {\xi^2 \over x R }\, \theta(x-\xi|y|) \ln(1+\sigma^2)\Bigg\}
\,,
\eea

\bea
&& W_4(q,k_f)^{(H)} = {g^4 M_N \over 8\pi^3}
\Bigg\{{2x^2 \over 3}-2u^2-{1\over 2} -\ln(1+4x^2) \nonumber \\
&& +{4x^2-3 \over 4x} \arctan 2x +\int_{-1}^1dy\, \bigg[ {1+u^2-x^2y^2
\over 1+s^2} \nonumber \\
&& +2xy\bigg({s\over 1+s^2}-\arctan s \bigg)+\ln(1+s^2)
\bigg]  \Bigg\}\,, 
\eea

\bea
&& W_4(q,k_f)^{(F)} =  {g^4 M_N \over 16\pi^3} \int_{-1}^1 dy
\int_0^u d\xi\, {\xi^2 \over x R }\Big[
\theta(x-\xi|y|) \nonumber \\
&& \times \theta(\xi-x)\ln(1+\sigma^2)+ \theta(x-\xi)
\ln(1+\sigma_x^2) \Big] \,, 
\eea
with $\sigma_x= \xi y +\sqrt{u^2-x^2+\xi^2y^2}$.

Similarly, the second-order contributions to $ U(q,k_f)+i\,W(q,k_f)$ for 
momenta outside the Fermi sphere $q>k_f$ are given by:
\bea
&& U_2(q,k_f)^{(H)} = {g^4 M_N \over 8\pi^3} \bigg\{
\arctan(u+x)-\arctan(x-u) \nonumber \\
&& -u+{x^2-u^2-1 \over 4x}
\ln{1+(u+x)^2 \over 1+(u-x)^2} \,,
\eea

\bea
 U_2(q,k_f)^{(F)} &=& {g^4 M_N \over 16\pi^3 x}
\int_{(x-u)/2}^{(u+x)/2} d\xi \, \Big[u^2-(2\xi-x)^2\Big] \nonumber \\
&&\times {\arctan 2\xi-\arctan \xi \over 1+2\xi^2} \,,
\eea

\bea
&& U_3(q,k_f)^{(H)} = {g^4 M_N \over 8\pi^4} \Bigg\{
\int_{y_{\rm min}}^1 dy\,\bigg\{\bigg[ uxy +{1\over 2} (u^2-x^2y^2) \nonumber \\
&&\times \ln{u+x y\over |u-x y|} \bigg] {\cal A}_y\bigg[{s^2 \over 1+s^2}\bigg]
+\int_{xy-s}^{s-x y} d\xi \bigg[2u \xi \nonumber \\
&& +(u^2-\xi^2) \ln{u+\xi \over u-\xi} \bigg] {x y +\xi \over [1+(xy +\xi)^2]^2}
\bigg\}\nonumber \\ && +{1\over x} \int_{-1}^1 dy\int_0^u d\xi \, {\xi^2 \sigma^2
\over 1+\sigma^2} \ln{x+\xi y \over x-\xi y} \Bigg\}\,,
\eea
with $y_{\rm min}=\sqrt{1-u^2/x^2}$ and the antisymmetrization
prescription ${\cal A}_y[f(y)]=f(y)- f(-y)$.

\bea
&& \hspace{-.16in} U_3(q,k_f)^{(F)} =  {g^4 M_N \over 16\pi^4}\Bigg\{
-\int_{y_{\rm min}}^1 dy\int_{y_{\rm min}}^1 dz\,  {\cal A}_y[\ln(1+s^2)]\nonumber \\
&& \hspace{-.16in} \times {\cal A}_z[\ln(1+t^2)] {\theta(y^2+z^2-1)
\over 4 \sqrt{y^2+z^2-1}}\, +\int_{-1}^1 dy \int_0^u
d\xi\, {\xi^2 \over x R } \nonumber \\ 
&& \hspace{-.16in} \times \ln(1+\sigma^2) \ln{x R +(x^2-\xi^2-1)
y \xi \over x R +(1+\xi^2-x^2) y \xi} \Bigg\}\,,
\eea

\bea
&&\hspace{-.1in} W_2(q,k_f)^{(H)} = {g^4 M_N \over 16\pi^3} \bigg\{
\ln{1+(u+x)^2\over 1+(u-x)^2} -{2u \over 3x}(3+2u^2) \nonumber \\ 
&& \hspace{-.1in}+{1+u^2-x^2 \over x} \Big[
\arctan(u+x)-\arctan(x-u)\Big] \bigg\} \,,
\eea

\bea
 W_2(q,k_f)^{(F)} &=& {g^4 M_N \over 32\pi^3x}
\int_{(x-u)/2}^{(u+x)/2} d\xi \,\Big[u^2-(2\xi-x)^2\Big] \nonumber \\
&& \times {\ln(1+4\xi^2) \over 1+2\xi^2} \,,
\eea

\bea 
&&\hspace{-.35in} W_{34}(q,k_f)^{(H)} = {g^4 M_N \over 16\pi^3}\Bigg\{
{1\over x}\bigg[ {u\over 2}+{4u^3 \over 3}-{1\over 4}(1+4u^2) \nonumber \\
&&\hspace{-.35in} \times \arctan 2u \bigg] + \theta(\sqrt{2}u-x) 
\int_{y_{\rm min}}^{u/x} dy\,(x^2y^2-u^2)  \nonumber \\
&&\hspace{-.35in} \times {\cal A}_y\bigg[{s^2 \over
1+s^2}\bigg] +\int_{y_{\rm min}}^1 dy\,{\cal A}_y\bigg[-\ln(1+s^2)
+{s^2 \over 1+s^2}  \nonumber \\
&&\hspace{-.35in} \times (1+u^2-x^2y^2)
 +2x y\bigg( \arctan s -{s\over 1+s^2}\bigg)\bigg]  \Bigg\}\,,
\eea

\bea
&& W_{34}(q,k_f)^{(F)} = {g^4 M_N \over 16\pi^3}
\Bigg\{ {\theta(\sqrt{2}u-x) \over 4\pi} \int_{y_{\rm min}}^1 dy\nonumber \\
&& \int_{y_{\rm min}}^1 dz\,{\theta(1-y^2-z^2) \over \sqrt{1-y^2-z^2}}
 {\cal A}_y[\ln(1+s^2)]{\cal A}_z[\ln(1+t^2)] \nonumber \\
&& -\int_{-1}^1 dy\int_0^u d\xi \,{\xi^2 \over x R} \ln(1+\sigma^2)
\Bigg\}\,,
\eea

Finally, we note that the total imaginary part $W(0,k_f)$ evaluated at zero 
nucleon-momentum ($q=0$) can even be written in closed analytical form:
\begin{eqnarray} 
&&\hspace{-.3in}W(0,k_f) = {g^4 M_N \over 16 \pi^3} \Bigg\{ {\pi^2 \over 12}
+{\rm Li}_2(-1-u^2) -{2u^2 \over 1+u^2} \nonumber \\ 
&& \hspace{-.2in}+ \bigg[ 2+\ln(2+u^2)-{1\over 2}  \ln(1+u^2)\bigg] \ln(1+u^2)\Bigg\} \,, 
\end{eqnarray}
where Li$_2(\dots)$ denotes the conventional dilogarithmic function. The
behavior of $W(0,k_f)$ at small densities is $k_f^4$.

\end{document}